# Local amplification of deep mining induced vibrations
## Part.2: Simulation of ground motion in a coal basin

Semblat J.F.[1], Lokmane N.[1,3], L. Driad-Lebeau[2], Bonnet G.[3]

[1] Université Paris-Est, LCPC, Dept of Soil and Rock Mechanics and Eng. Geology, 58 bd Lefebvre, 75015 Paris, France, semblat@lcpc.fr ,
[2] Institut National de l'Environnement Industriel et des Risques (INERIS) – Ecole des Mines, Parc Saurupt 54042 Nancy, France, Lynda.Driad-Lebeau@ineris.fr,
[3] Université Paris-Est, Laboratoire de Modélisation et Simulation Multi-Echelle LMSME (CNRS UMR 8208), France, guy.bonnet@univ-paris-est.fr

**Abstract**
This work investigates the impact of deep coal mining induced vibrations on surface constructions using numerical tools. An experimental study of the geological site amplification and of its influence on mining induced vibrations has already been published in a previous paper (Part 1: Experimental evidence for site effects in a coal basin). Measurements have shown the existence of an amplification area in the southern part of the basin where drilling data have shown the presence of particularly fractured and soft stratigraphic units. The present study, using the Boundary Element Method (BEM) in the frequency domain, first investigates canonical geological structures in order to get general results for various sites. The amplification level at the surface is given as a function of the shape of the basin and of the velocity contrast with the bedrock. Next, the particular coal basin previously studied experimentally (Driad-Lebeau et al., 2009) is modeled numerically by BEM. The amplification phenomena characterized numerically for the induced vibrations are found to be compatible with the experimental findings: amplification level, frequency range, location. Finally, the whole work was necessary to fully assess the propagation and amplification of mine induced vibrations. The numerical results quantifying amplification can also be used to study other coal basins or various types of alluvial sites.

## 1 Mine induced vibrations

As shown in Driad-Lebeau et al. (2009), mining operations may induce a redistribution of the stress field based on the mechanical behavior of the rockmass. This can lead to a substantial microseismic activity (Ben Slimane et al, 1990; Linkov et al, 1997; Kanelo et al; 1999; Senfaute et al, 1997; Senfaute et al; 2001; Driad-Lebeau et al, 2005). The rupture process generates elastic waves, which are propagated through the geological structure up to the free surface. Seismic monitoring was thus performed in numerous mines (Driad-Lebeau et al., 2009). In recent years, the impact of mine induced vibrations on surface constructions (i.e. houses or buildings located close to a mine) has been studied. This type of dynamic loading is different from seismic excitations coming from natural earthquakes (return period, amplitude, frequency range, etc).

Detailed studies were carried out for a coal basin in the framework of a French research program called "SisMine" initiated by INERIS and sponsored by the French collieries (Driad-Lebeau et al., 2009). LCPC and University Paris-Est-Marne la Vallée were associated to this research program in order to develop a numerical methodology aimed at simulating the impact on surface constructions of weak amplitude vibrations. The SisMine research program is subdivided into three parts, each one devoted to specific goal: Part1/ Experimental estimation of site effects in the coal basin (Driad-Lebeau et al., 2009); Part2/ Numerical estimation of site effects in the coal basin and comparison with experimental results (present paper); Part3/ Impact of deep mining vibration on surface constructions – Numerical approach.

This paper investigates numerically the propagation and amplification of mine induced vibrations in coal basins. It consists in a general study for various basin geometries (canonical basins) and detailed analyses for the Gardanne coal basin (Provence, France). Comparisons with experimental results from the field are also proposed.

## 2 Experimental analysis in the field

### 2.1 Site description

The Gardanne basin is located between Aix-en-Provence and Marseille (South of France) several kilometers westward from the city of Gardanne (latitude: 43° 27' 16" North and longitude : 5° 28' 34" East). It overlays a coal field which forms the eastern part of the arc basin and constitutes an E-W oriented geological unit. The general tectonic features and geological setting of the basin are quite simple (Fig. 1). The Gardanne basin is composed of a fluvio-lacustrine of the upper Cretaceous and the Eocene overlaying a substratum of the Jurassic (or lower cretaceous). The stratigraphy sequence consists mainly in marls, limestones and sandstones of the Valdonnian together with limestones of the Fuvelian (hard and brittle) with intermediate lignites (Fig. 1). The shallower sequences are represented by clay-sandy limestones of the Rognacian and Begudian. The presence on the surface of marine abrasion and molasses deposits has marked the influence of a sedimentary episode of the Miocene.

Among the eight coal levels having been exploited since the Middle Ages, the last seam mined until the closing of the mine (2003) was the so called "Grande Mine". That is the most significant layer (2.5 m thick) located at depths ranging from 1000 m to 1400 m. The coal layers were worked with a long wall caving method that uses two roadways and extracts coal along a straight front having a large longitudinal extension. The stoping area close to the face is kept open to provide a security zone for the staff and the mining equipment.

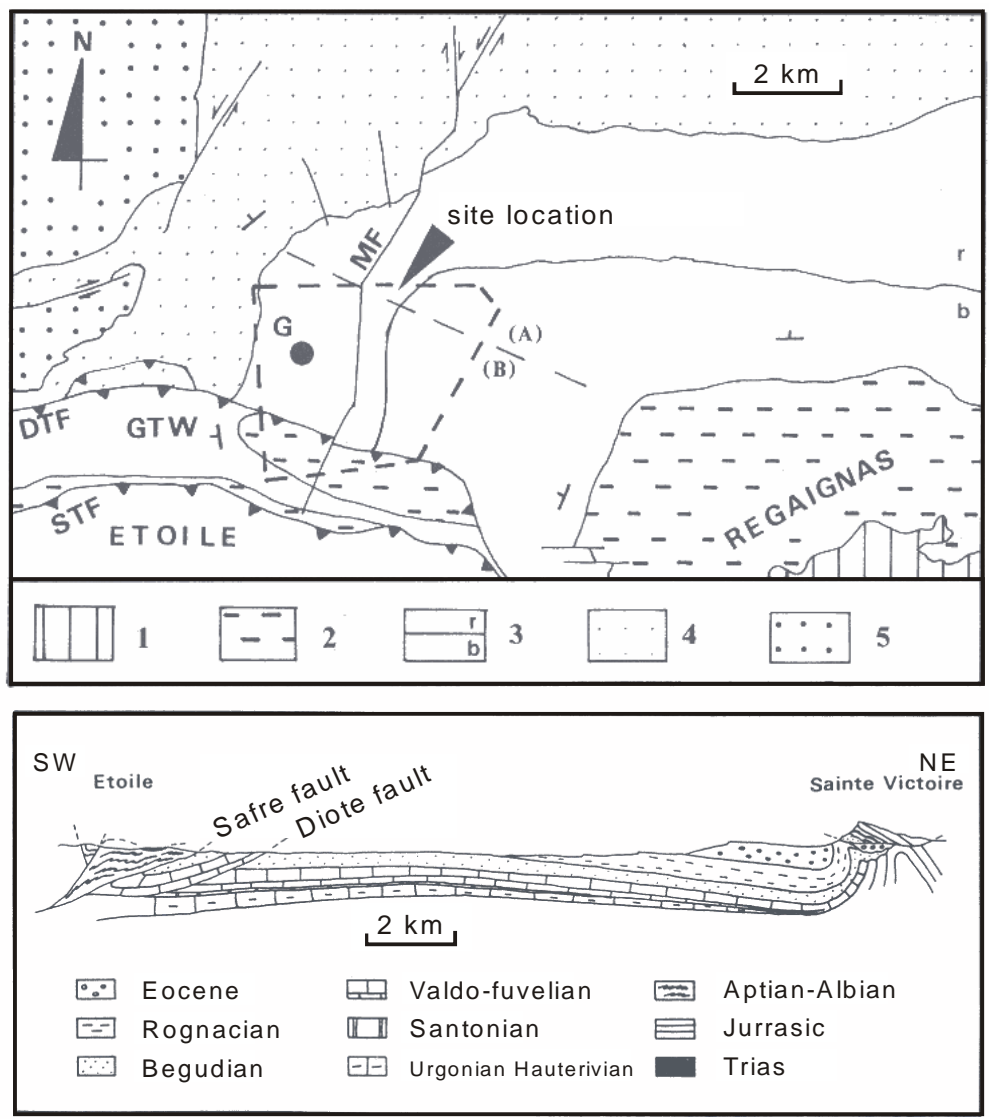

**Figure 1. Top:** Geological setting and location of Gardanne colliery (1: Upper Jurassic; 2: Campanian; 3: Begudian and Rognacian (3b,3r); 4: Eocene; 5: Oligocene). The coalfield is located in the Campagnian limestones.
**Bottom:** Geological cross-section of the Gardanne basin *[after Guieu, 1968, Durand et al., 1980; Tempier and Durand, 1981]*

### 2.2 Experimental results on induced vibrations

The seismic events induced by mining exploitation were recorded by using the mobile network described in (Driad-Lebeau et al., 2009). These data have been processed in the frequency domain and all events of magnitude greater than 2.5 have been considered (mine depth is approximately 1 km). Such events constitute the vibrations of interest in terms of impact on surface constructions. In figure 2, H/V spectral ratios from mine induced vibrations recordings were computed for 10-instrumented sites (8 residences and 2 free-surface sites). They are plotted as average H/V spectral ratios plus/minus one standard deviation.

As shown in Figure 2, the H/V spectral ratios highlight significant variations in resonances and amplitude peaks (evidencing amplification) at the investigated sites. Spectral ratios above 8 are found in the frequency band 3-8 Hz at sites FOU, LER, LAG and MON and NAY (see (Driad-Lebeau et al., 2009) for these various locations). The sites HEN and NAY, where outcrops are mainly marl-limestone, present a weak resonance (amplitude of nearly 3-4) at 3-6 Hz. This observation is coherent with the geological setting where the limestone dominates. In this particular case, the amplification effect is not very significant. It is interesting to note the response of the site MON, which presents a broad resonance at 4.5 Hz with an amplitude of 7. Indeed, according to the geology (limestone-marls), the H/V ratio would be expected close to that of the site HEN. It suggests that the observed amplitude could be related to a topographic effect. Indeed, the corresponding house is located on the slope of a hill, which culminates at 210 meters.

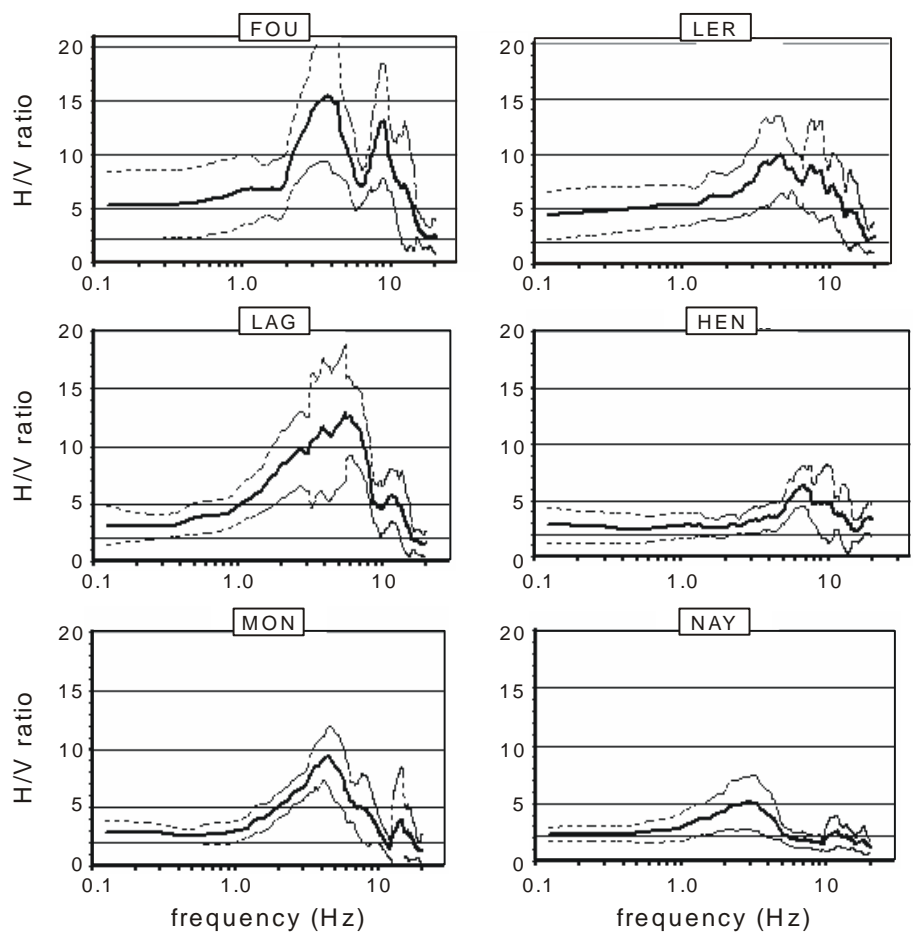

**Figure 2**: H/V spectral ratios from mining induced seismic data. Thick line: average H/V ratios for each recording location, dotted line: plus/minus one standard deviation.

## 3 Modeling wave propagation in soils

### 3.1 Numerical methods for wave propagation

To analyze wave propagation (seismic waves, vibrations, etc) in 2D or 3D geological structures, various numerical methods are available:

- the finite difference method is accurate in elastodynamics but free surface or interface conditions has to be carefully considered (Moczo 2002, Virieux 1986),
- the finite element method is efficient to deal with complex geometries and numerous heterogeneities (even for inelastic constitutive models (Bonilla, 2000)) but has several drawbacks such as numerical dispersion (error in terms of phase velocity) and numerical damping (Hughes 2008, Ihlenburg 1995, Semblat 2000a, Semblat and Pecker 2009) and (consequently) numerical cost in 3D elastodynamics,
- the spectral element method has been increasingly considered to analyse 2D/3D wave propagation in linear media with a good accuracy due to its spectral convergence properties (Chaljub 2007, Faccioli et al. 1996, Komatitsch et al. 1998),
- the boundary element method allows a very good description of the radiation conditions but is preferably dedicated to weak heterogeneities and linear constitutive models (Beskos 1997, Bonnet 1999, Dangla et al. 2005, Semblat et al. 2000b). Recent developments have been proposed to reduce the computational cost of the method especially in the high frequency range (Chaillat et al. 2008, 2009, Fujiwara 2000),
- the Aki-Larner method which takes advantage of the frequency-wavenumber decomposition (Aki 1970, Bouchon 1989),
- the scaled boundary finite element method which is a kind of solution-less boundary element method (Wolf 2003),
- other methods for simple geometries such as series expansions of wave functions (Sanchez-Sesma 1983).

Furthermore, when dealing with wave propagation in unbounded domains, many of these numerical methods require absorbing boundary conditions to avoid spurious reflections (Chaljub et al. 2007, Semblat & Pecker 2009). It is for instance possible to couple FEM and BEM (Aochi 2005, Bonnet 1999) allowing an accurate description of the near field (FEM model including complex geometries, numerous heterogeneities and nonlinear constitutive laws) and a reliable estimation of the far-field (BEM involving accurate radiation conditions).

### 3.2 The boundary element method

The main advantage of the boundary element method is to avoid artificial truncation of the domain in the case of an infinite medium. For dynamic problems, this truncation leads to artificial wave reflections giving a numerical error in the solution. The boundary element method can be divided into two main stages (Bonnet 1999):

- solution of the boundary integral equation giving displacements and stresses along the boundary of the domain,
- a posteriori computation for all points inside the domain using an integral representation formula.

The boundary element method arises from the application of Maxwell-Betti reciprocity theorem leading to the expression of the displacement field inside the domain $\Omega$ from the displacements and stresses along the boundary $\partial\Omega$ of the domain (Bonnet 1999).

### 3.3 Elastodynamics

We consider an elastic, homogeneous and isotropic solid of volume $\Omega$ and external surface $\partial\Omega$. Within this medium, the equation of motion can be written under the following form:

$$(\lambda + 2\mu)\,grad\,(div\,\mathbf{u}) - \mu\,rot\,(rot\,\mathbf{u}) + \rho\mathbf{f} = \rho\ddot{\mathbf{u}} \tag{1}$$

where $\lambda$ and $\mu$ are the Lamé coefficients, $\mathbf{u}$ the displacement field, $\rho$ the mass density and $\mathbf{f}$ a force density.

By using the Fourier transform, the problem can be studied in the frequency domain, for each circular frequency $\omega$. The equation of motion for a steady state ($\mathbf{u}(x)$, $\sigma(x)$) can then be written as follows:

$$(\lambda + 2\mu)\,grad\,(div\,\mathbf{u}(x)) - \mu\,rot\,(rot\,\mathbf{u}(x)) + \rho\mathbf{f}(x) + \rho\omega^2\mathbf{u}(x) = 0 \tag{2}$$

This equation is written in the framework of linear elasticity but, since the analysis is performed in the frequency domain, damped mechanical properties may be considered through the complex modulus of the medium (Semblat and Pecker 2009).

### 3.4 Integral representation

For steady solutions of harmonic problems, the reciprocity theorem between two elastodynamic states comprising displacement fields and stress fields ($\mathbf{u}(x)$, $\sigma(x)$) in equilibrium with body forces $\mathbf{f}(x)$ and ($\mathbf{u}'(x)$, $\sigma'(x)$) in equilibrium with body forces $\mathbf{f}'(x)$ takes the following form:

$$\int_{\partial\Omega}\mathbf{t}^{(n)}(x)\mathbf{u}'(x)ds(x) + \int_{\Omega}\rho\mathbf{f}(x)\mathbf{u}'(x)dv(x) =$$
$$\int_{\partial\Omega}\mathbf{t}'^{(n)}(x)\mathbf{u}(x)ds(x) + \int_{\Omega}\rho\mathbf{f}'(x)\mathbf{u}(x)dv(x) \tag{3}$$

The integral formulation is obtained through the application of the reciprocity theorem between the elastodynamic state ($\mathbf{u}(x)$, $\sigma(x)$) and the fundamental solutions of a reference problem called Green kernels. The reference problem generally corresponds to the infinite full space case in which a volumic concentrated force at point $y$ acts in direction $\mathbf{e}$. In the harmonic case, the Green kernel of the infinite medium corresponds to a volumic force field such as:

$$\rho\mathbf{f}'(x) = \delta(x - y)\mathbf{e} \tag{4}$$

In this article, the model involves the Green functions of an infinite medium (Bonnet 1999) or semi-infinite medium (in the case of SH-waves). The Green kernel is denoted $U_{ij}^{\omega}(x,y)$ and characterizes the complex displacement in direction $j$ at point $x$ due to a unit (time harmonic) force concentrated at point $y$ along direction $i$. The corresponding traction on a surface of normal vector $n(x)$ is denoted by $T_{ij}^{(n)\omega}(x,y)$.

The application of the reciprocity theorem between the elastodynamic state ($\mathbf{u}(x)$, $\sigma(x)$) and that defined by the Green kernel $U_{ij}^{\omega}(x,y)$ gives the following integral representation:

$$I_{\Omega}(y)u_i(y) = \int_{\partial\Omega}\left(U_{ij}^{\omega}(x,y)t_j^{(n)}(x) - T_{ij}^{(n)\omega}(x,y)u_j(x)\right)ds(x)$$
$$+ \int_{\Omega}\rho U_{ij}^{\omega}(x,y)f_j(x)dv(x) \tag{5}$$

where $I_{\Omega}(y)$ is 1 when $y$ is inside $\Omega$ and 0 when it is outside $\Omega$.

Numerical solution of equation (5) can be performed by collocation method or by an integral variational approach (Bonnet 1999).

### 3.5 Regularization and discretization of the problem

The integral representation defined by equation (5) is generally not valid for $x \in \partial \Omega$. The formulation of the boundary integral equation along $\partial \Omega$ is then not very easy to obtain as the Green kernels have singular values when $x \in \partial \Omega$. It is then necessary to regularize expression (5) to write the boundary integral equation (Beskos 1997, Bonnet 1999, Dangla et al. 2005).

The problems presented in this article are analyzed in two dimensions (plane or anti-plane strains). Two-noded boundary elements are chosen and the element size corresponds to one-tenth of the minimum wavelength. Two dimensional Green kernels of the infinite space are written using Hankel's functions (Bonnet 1999). The regularized solution of equation (5) is estimated by classical boundary finite elements discretization and then by collocation method, that is application of the integral equation at each node of the mesh.

## 4 Wave amplification in simple alluvial structures

Many different authors have studied the propagation and amplification of seismic waves in alluvial structures (Bard 1995, Bouchon 1973, Chavez-Garcia et al. 2000, Sanchez-Sesma et al. 2000, Semblat et al. 2000b, 2005). In such geological structures, the seismic motion may be amplified due to the velocity contrast between the various layers but also due to the limited geometrical extent of the basin (trapped surface waves). For mine induced vibrations, this phenomenon may also occur and significantly modify the ground motion at the free surface. We will thus analyze ground motion amplification, in a first step for simplified geological structures and next for the actual profile of the Gardanne coal basin.

### 4.1 Preliminary analysis for various geological deposits

Few geotechnical data are available for the Gardanne coal basin and we thus performed a parametric study making our results useful for other sites. Various geometries have been chosen for the deposit with variable mechanical properties. The incident wavefield is a plane vertical SH wave. The numerical simulations involve the Boundary Element Method (FEM-BEM code CESAR-LCPC (Humbert et al. 2005)).

The amplification of seismic waves in alluvial deposits is strongly influenced by the mechanical properties of the latter. Indeed the velocity contrast between soil layers governs the ground motion at the free surface. The geometry of the deposit is also an important factor. It may be characterized by its mean depth or in a more detailed way for alluvial basins. Due to the lateral heterogeneities, the seismic waves are trapped in the basin, leading to a large motion amplification. In the one-dimensional case (horizontal layers), a close-form solution can be obtained for the amplification factor of the ground motion (Semblat and Pecker 2009). Conversely, when the lateral heterogeneities are strong, 2D or 3D wave propagation must be considered.

### 4.2 Horizontal layering

We first consider the case of a simplified deposit involving a single horizontal layer. The layer depth is estimated from the Simiane 2 (SI2) borehole giving an approximate depth but no detailed information about the layer geometry (Driad-Lebeau et al. 2009). The fundamental solution for the half space is considered and a vertically incident plane wave is propagated in the bedrock. The simplified geometry is defined as follows:

- Layer depth H=15m and width L=3000m;
- Layer and bedrock properties (subscript *L* and *b* respectively):
  - Case 1: bedrock/layer velocity contrast $V_{Sb}/V_{SL}$=12;
  - Case 2: bedrock/layer velocity contrast $V_{Sb}/V_{SL}$=4.

As displayed in Figure 3, the maximum amplification factor is 16.75 in case 1 (reached at f=1.8Hz) and 5.0 in case 2 (at f=4.4Hz). As shown by these results (also see closed-form solutions in (Semblat and Pecker 2009)), the ground motion amplification is strongly influenced by the velocity contrast between the soil layer and the bedrock.

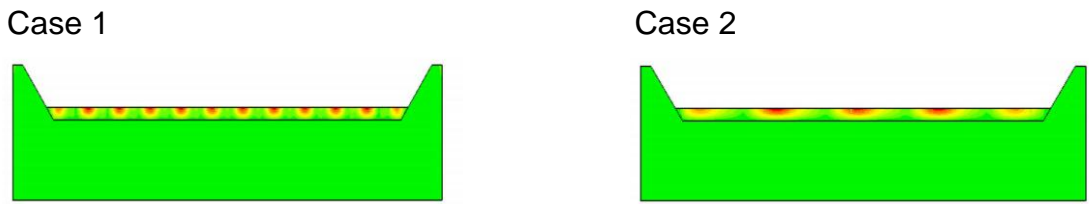

**Figure 3:** Ground motion amplification in the alluvial deposit. Maximum amplification is obtained in the red zone. Case 1 (left): maximum amplification 16.75 at f=1.8Hz; Case 2 (right): maximum amplification 5.0 at f=4.4Hz.

When comparing these results to the 1D closed-form solutions, the former are approximately 40% larger than the latter. The geometrical extent of the deposit also has a strong influence on the ground motion amplification. In the following, we will thus analyze the influence of the basin geometry on the motion amplification.

### 4.3 Influence of the basin geometry

#### 4.3.1 Variable shape ratio

The influence of the basin geometry is assessed by considering elliptical basins with different geometrical extents (i.e. shape ratios). As shown in Figure 4, the basins are characterized by their half-width *L*, their depth *H* and their shear wave velocity $V_{S1}$ ($V_{S2}$>$V_{S1}$ being the shear velocity velocity in the bedrock). Various geometries are considered: narrow basins (*L*<*H*) as well as large ones (*L*>*H*).

In order to use these results for various alluvial sites, the parametric study is performed considering such dimensionless parameters as: the amplification factor *A*, the horizontal shape ratio $\kappa_h = L/H$ ($\kappa_h$=1 for the circular geometry), the velocity ratio $\chi = V_{S2}/V_{S1}$ and the dimensionless frequency $\eta_v = H/\lambda$ (ratio between basin depth and wavelength).

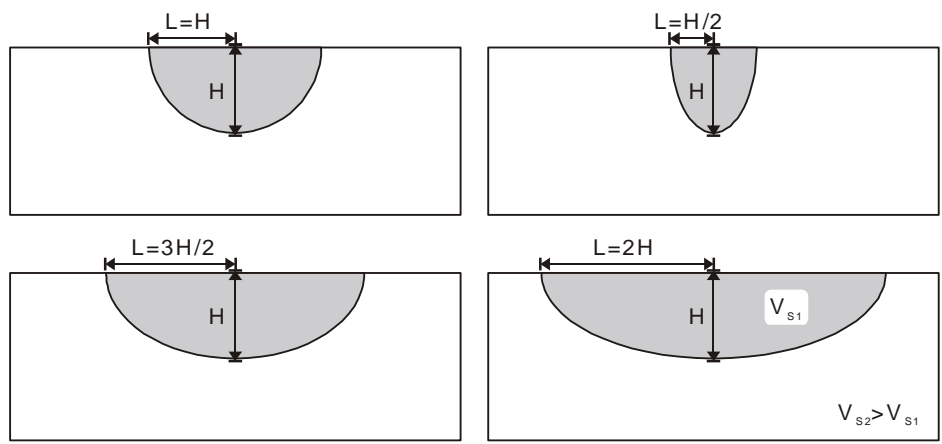

**Figure 4:** Elliptical basins of variable shape ratio $\kappa_h = L/H$.

*4.3.2 Comparison between the 1D case and the circular basin*
Considering an alluvial deposit of constant thickness overlying an elastic bedrock, the transfer function of the ground motion across the soil layer may be derived as a closed-form solution (Semblat and Pecker 2009). For a plane SH wave, the modulus of the transfer function is:

$$\left|T^*_{1,2}(\omega)\right| = \left[\cos^2 k_{z1}H + \bar{\chi}^2 \sin^2 k_{z1}H\right]^{-1/2} \quad (6)$$

where $k_{z1} = \dfrac{\omega \cos \alpha_1}{V_{S1}}$ is the horizontal wavenumber and $\bar{\chi} = \sqrt{\dfrac{\rho_1 \mu_1}{\rho_2 \mu_2} \dfrac{\cos \alpha_1}{\cos \alpha_2}}$ with $\omega$ the frequency, $\alpha_1$ and $\alpha_2$ the angles between the direction of propagation of the wave and the vertical axis in the layer and in the bedrock respectively ($\alpha_1$ is estimated from $\alpha_2$ (Semblat and Pecker 2009)). $H$ is the layer depth, $\mu_1$, $\mu_2$ are the shear moduli and $\rho_1$, $\rho_2$ the mass densities (indices 1 and 2 for the layer and bedrock resp.).
Equation (6) corresponds to the amplification factor of the ground motion: the amplitude at the top of the layer is divided by the so-called outcrop motion (amplitude at the surface of the bedrock without the alluvial deposit).

From Equation (6), the dimensionless frequency giving the maximum amplification is found to be $\eta_v = 0.25$ (i.e. quarter-wavelength resonance) whereas, for the circular basin, the maximum amplification is reached at $\eta_v = 0.35$ (Semblat and Pecker 2009). The different is due to the 2D basin effects (lateral heterogeneities). We will now investigate the influence of the basin shape on the amplification level.

## 5 Variable basin shape: parametric study

For a plane SH-wave, various elliptical basins are considered (Figure 4). Their horizontal shape ratios, $\kappa_h = L/H$, are chosen as $\kappa_h = 0.5; 1; 2; 3; 4; 5$ and $6$ (the basin depth being constant: $H=25$m). Different velocity ratios were also chosen: $\chi=2$ to $8$. From all these models, the maximum motion amplification and the related frequency were computed. The results are plotted in Figure 5 as an abacus: solid lines correspond to fixed shape ratios $\kappa_h$ and dotted lines to constant velocity ratios $\chi$. The main conclusions are the following:

- For a constant velocity ratio and shape ratios larger than 1, the maximum amplification and the related frequency decrease when increasing shape ratio and the results are becoming closer to the 1D case.
- For narrow basins (small shape ratios), the 2D results are far from the 1D analysis (strong 2D effects).
- For a constant shape ratio, when the velocity ratio increases, the maximum amplification increases and the related frequency decreases.

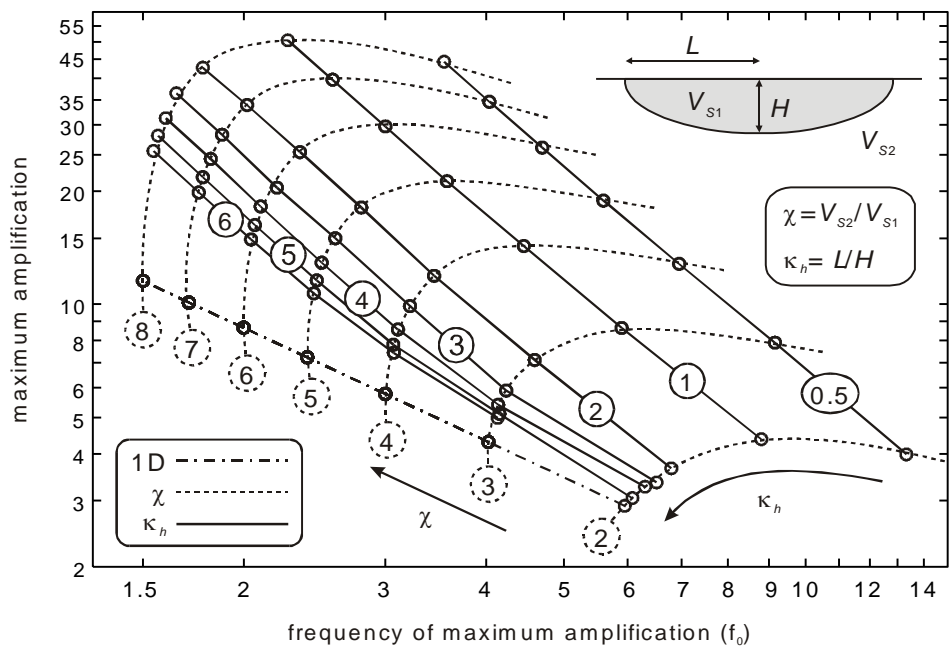

**Figure 5:** Maximum amplification and related frequencies (in Hz) for variable shape ratios $\kappa_h = L/H$ and velocity ratios $\chi = V_{S2}/V_{S1}$.

From Figure 5, it is thus possible to estimate the maximum ground motion amplification and the related frequency for various types of alluvial basins. For instance, if we consider $\kappa_h$=1.5 and $\chi$=3.5, the maximum amplification is above 10 and the related frequency around 4.5.

Our numerical results for an elliptical basin were compared to Bard and Bouchon's results (1985) for sinusoidal basins in terms of fundamental frequencies. Bard and Bouchon (1985) proposed the following empirical law:

$$f_{2D} = \left(\frac{V_{S1}}{4H}\right)\sqrt{1+\kappa_v^2} \qquad (7)$$

where $\kappa_v = 1/\kappa_h = H/L$ is the vertical shape ratio defined by Bard and Bouchon, $L$ is the basin half width, $H$: the basin depth and $V_{S1}$ the velocity in the basin.

When compared to Bard and Bouchon's results, our frequencies of maximum amplification have similar variations with respect to the shape ratio. 2D effects are found to be strong for narrow basins whereas large basins lead to amplification levels close to the 1D case.

The ground motion amplifications are computed in the frequency domain (time harmonic signal). Time domain computations will be proposed in the following.

## 6 Numerical analysis for the Gardanne coal basin

### 6.1 Coal basin profile

For the Gardanne coal basin (Fig. 1), a North-South profile has been defined (Fig. 6) along which four drillings have been performed ($F_1$, $F_2$, $F_3$, $F_4$). A Boundary Element model (FEM-BEM code CESAR-LCPC) has been prepared from this profile (Figure 6, bottom). From the control points $F_1$ and $F_3$, the coal basin has been defined using an elliptical curve and the maximum depth found at point $F_2$.

An alluvial deposit of finite extent (domain 1) is thus defined (Figure 6, bottom). The BEM computation is performed in the frequency domain by considering plane incident SH waves. The influence of the radiation pattern of the source may be strong (Crouch 1980, Rudnicki 1983, Semblat & Pecker 2009) and a detailed analysis of this issue should be considered. Nevertheless, since the recordings correspond to various sources (averaged results), the influence of the source location and type was disregarded and we have only considered plane wave excitation in the simulations. In our BEM mesh, the smallest Boundary Element size is chosen as 3 m along the free surface, allowing computations up to 20 Hz. Time domain solutions are computed afterwards from frequency responses by using inverse Fourier transform.

The mechanical features of the basin and the bedrock are the following: shear wave velocity in the basin (domain 1) $V_{S1}$=250 m/s, shear wave velocity in the bedrock (domain 2) $V_{S2}$=1200 m/s. Maximum depth of the surficial layer 11 m.

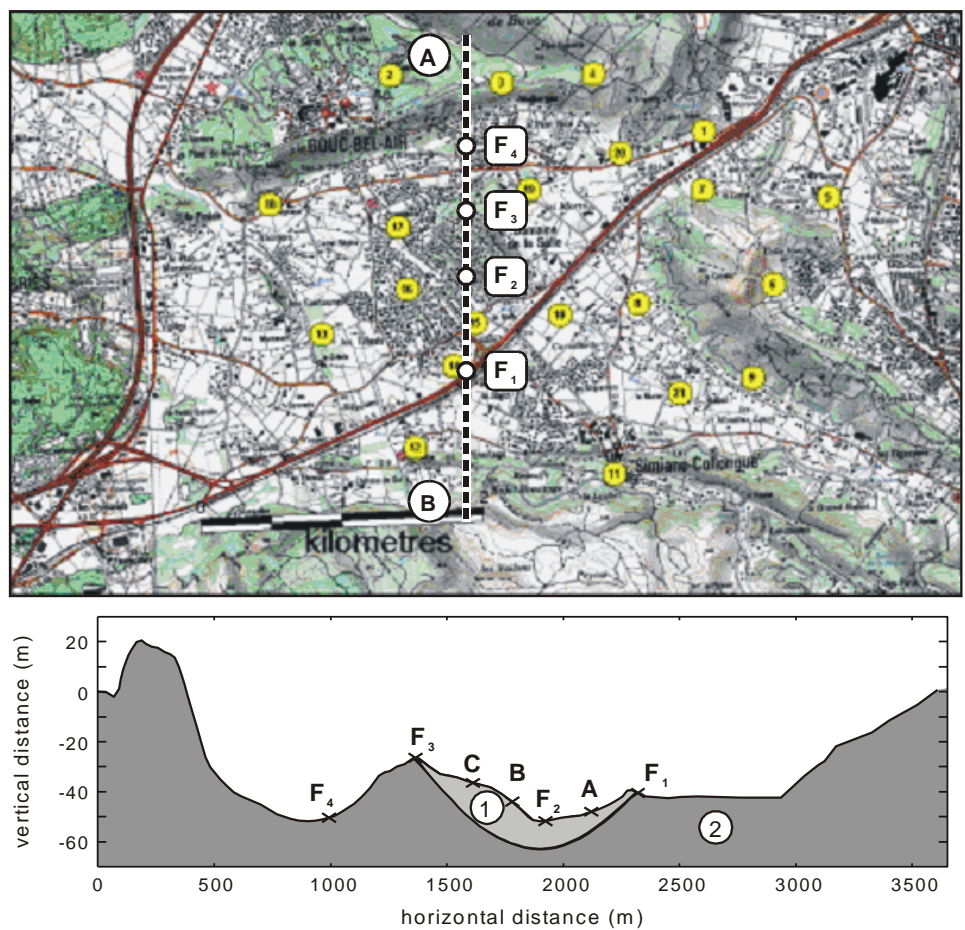

**Figure 6:** North-South profile and location of the four drillings $F_1$, $F_2$, $F_3$ and $F_4$ (top) and 2D geological model considered for the Boundary Element Method (bottom).

## 6.2 Amplification factors estimated numerically

The amplification level corresponds to the spectral ratios between the ground motion at the free surface and the outcrop motion (i.e. motion at the top of the bedrock when there is no alluvial deposit). The spectral amplification along the North-South profile is plotted in Figure 7 as a function of distance (between $F_3$ and $F_1$) and frequency (1.5 to 20Hz).

The largest amplification ($A$=14) is reached at f=4.6Hz and d=1500 m, at f=4Hz and d=1550m and also at f=3.9Hz and d=1580m or d=1720m. For the same mechanical and geometrical features, a 1D model leads to a fundamental frequency f=5.7Hz and a spectral amplification $A_0$=7 (Equation 6). Due to basin edge effects (trapped surface waves), the 2D amplification computed by the BEM is larger than the 1D amplification.

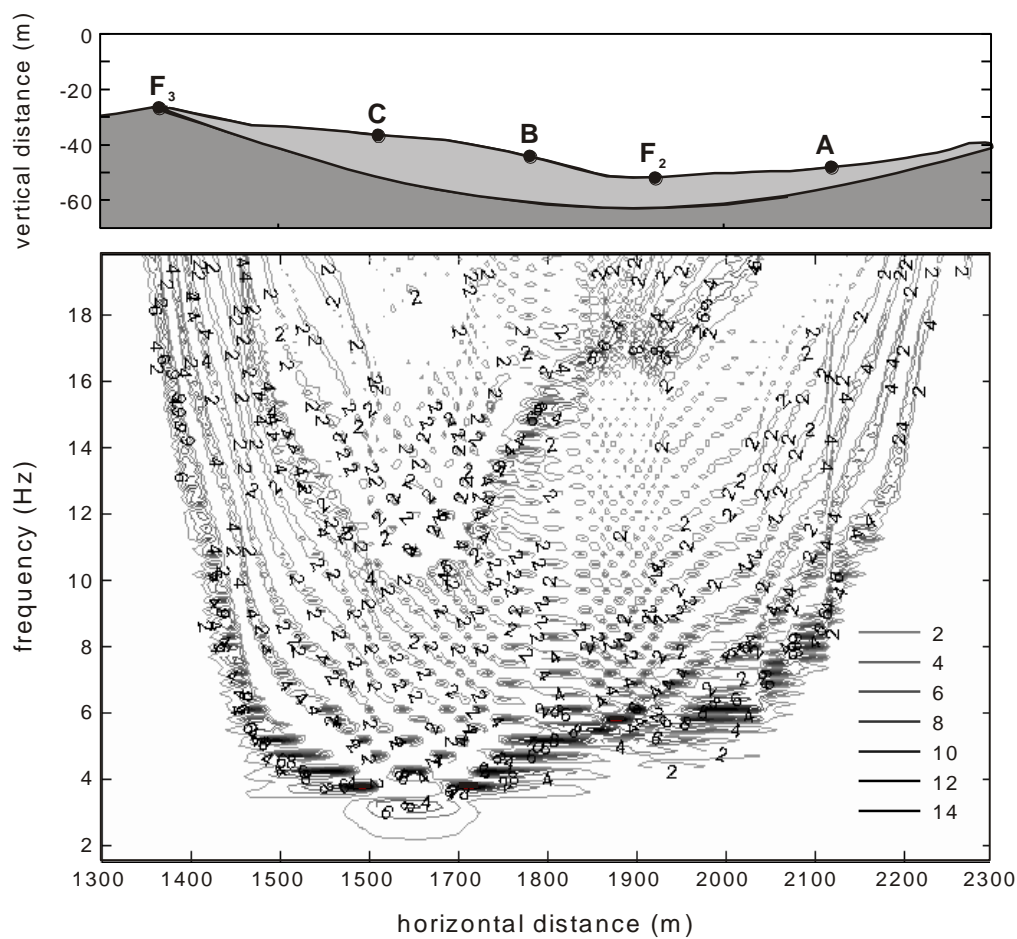

**Figure 7:** Amplification factor in the Gardanne coal basin as a function of position and frequency.

In Figure 6 (bottom), five points were identified along the basin surface: $F_1$, $d_1$=2314.8m; A, $d_A$=2107.2m; $F_2$, $d_2$=1902.4m; B, $d_B$=1691.2m; C, $d_C$=1499.3m. For these points, the transfer functions (outcrop motion) are displayed in Figure 8 and lead to the following results:

- At point $F_1$, a low motion amplification is obtained: around $A$=1.4 for $f$=5.5Hz. For the $F_1$ drilling location, the depth of the surficial layer is nearly zero and the amplification is thus small;

- At point A, the spectral amplification is nearly 12.5 at $f$=7.5Hz;
- The largest amplification (14) is found above drilling $F_2$ at frequency $f$=5.62Hz (this frequency value being close to the 1D fundamental frequency);
- At point B, the frequency of maximum amplification is $f$=3.8Hz and the amplification level is around 9.2;
- Finally, at point C, the maximum amplification (8.8) is reached at $f$=8Hz.

The maximum amplification derived from the 2D BEM model thus reaches 14 and is larger than the one estimated from the 1D solution (i.e. 7.7). This difference is mainly due to basin edge effects leading to trapped surface waves (Semblat et al. 2005). Finally, the order of magnitude of spectral amplifications estimated numerically is in the same range as the one of spectral ratios obtained from the recordings (Figure 2).

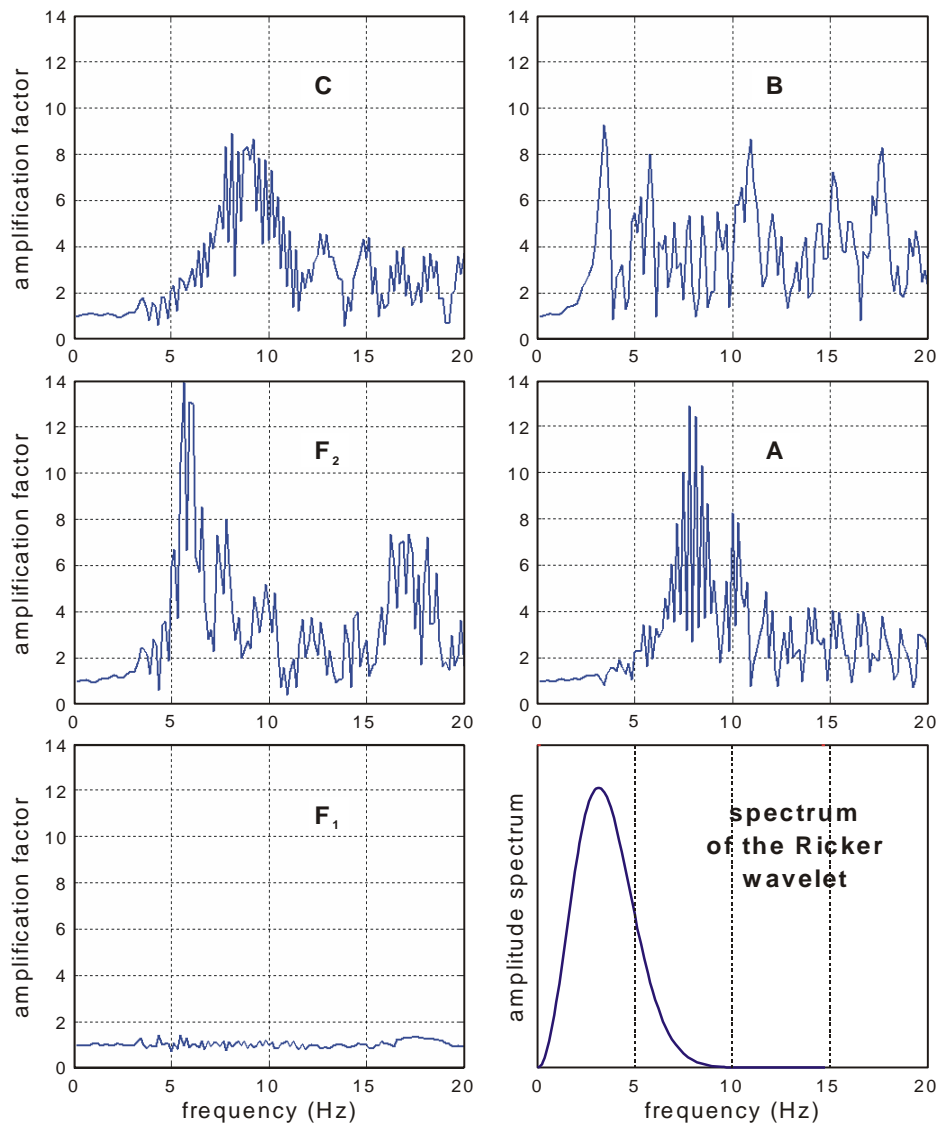

**Figure 8:** Transfer function (outcrop motion) for five different points along the deposit ($F_1$; A; $F_2$; B; C, see Figure 6) and spectrum of the Ricker wavelet used hereafter for the time-domain analysis (bottom right).

### 6.3 Amplification of synthetic wavelets

To investigate the motion amplification in the time domain, a 2$^{nd}$ order Ricker wavelet is now considered for the incident motion. The 2$^{nd}$ order Ricker wavelet corresponds to the 2$^{nd}$ derivative of a Gaussian (Semblat and Pecker 2009) and it is well localized both in time and frequency. The fundamental period of the Ricker wavelet is $t_p$=0.32s related to a fundamental frequency $f_p$=3.12Hz. Its amplitude spectrum is given in Figure 8 (bottom right) with the transfer functions.

Using the Fourier transform of the Ricker wavelet and the transfer functions at the five different points along the free surface (Figure 8), the ground motion at each point is determined in the time domain.

The time domain response is normalized by the amplitude of the incident Ricker wavelet $\bar{u} = u/u_0$ and plotted in Figure 9. The lowest amplification is found at point $F_1$ where only the free surface effect is observed ($\bar{u} = u/u_0 = 2$ due to the reflection). The largest amplification is reached at point $F_2$ ($\bar{u}$ =4.09) and point B ($\bar{u}$ =4.71) (Figure 9).

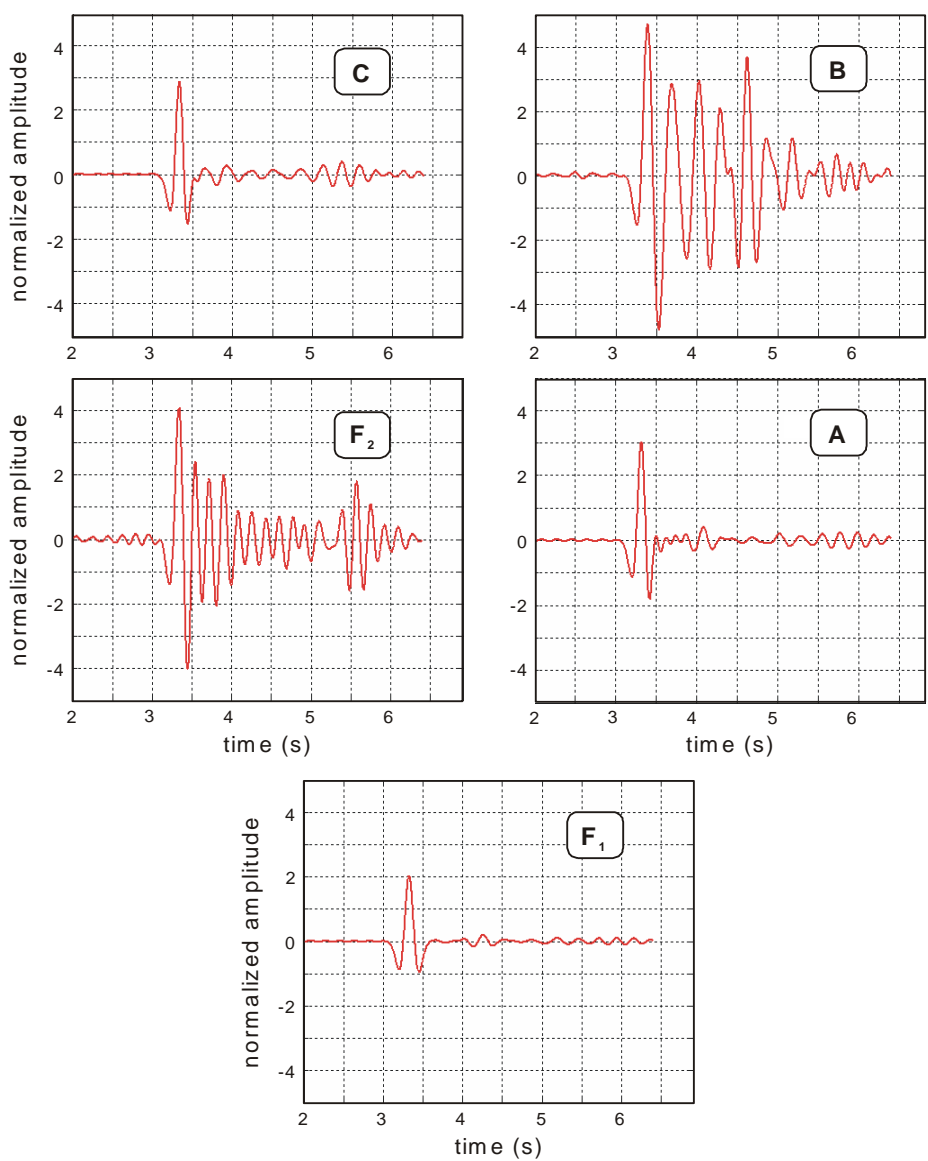

**Figure 9:** Ground motion at the free surface due to an incident Ricker wavelet.

When comparing the transfer functions at points $F_2$ and B at the fundamental frequency of the Ricker wavelet, $f_p$=3.12 Hz (Figure 8), the spectral amplification at point B is larger than that at point $F_2$. The time domain amplification would have been larger at point $F_2$ for a Ricker wavelet with a larger fundamental frequency. At both points, a strong increase of the motion duration is also observed (Figure 9). It is not the case at points A et C where the amplification is low and the motion duration is hardly amplified.

### 6.4 Amplification from actual signals

*6.4.1 Reference motion at the basin edge (point $F_1$)*

To compute the ground motion using actual recordings, it is mandatory to get a reference outcrop motion in order to combine it with the transfer functions at each point along the basin. However, since it is often difficult to have a good outcrop reference site, the reference motion may also be obtained by deconvoluting a recorded surface motion using the transfer function at this point (it must have been computed in an accurate way). From the reference signal, the ground motion may then be computed all along the alluvial deposit.

As shown in Figure 8, the transfer function at point $F_1$ is very flat and very close to 1 (low amplification). The ground motion at this point may thus be considered as the reference motion in order to compute the surface motion inside the basin. Since the BEM computations involve SH waves (*y*-polarization normal to the model plane), the acceleration component $a_y$ is chosen. The reference outcrop acceleration $a_y$ at point $F_1$ is displayed in Figure 10. It has been weighted by a Hamming window and the peak acceleration is 0.155 m/s$^2$.

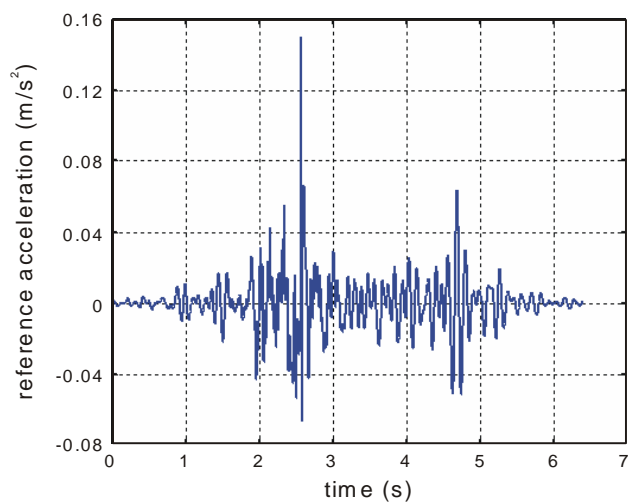

**Figure 10:** Reference outcrop acceleration $a_y$ at the basin edge (point $F_1$).

*6.4.2 Recorded motion at the basin center (point $F_2$)*

The three components of the ground motion recorded at the center of the basin (point $F_2$) are displayed in Figure 11 in terms of particle velocity (left) and acceleration. The amplitude of the particle velocities is small compared to seismic motions (PGV=10mm/s). The largest acceleration is reached along *y*: $a_y$=0.407m/s$^2$. In the following, the ground motion will be computed for different points along the basin.

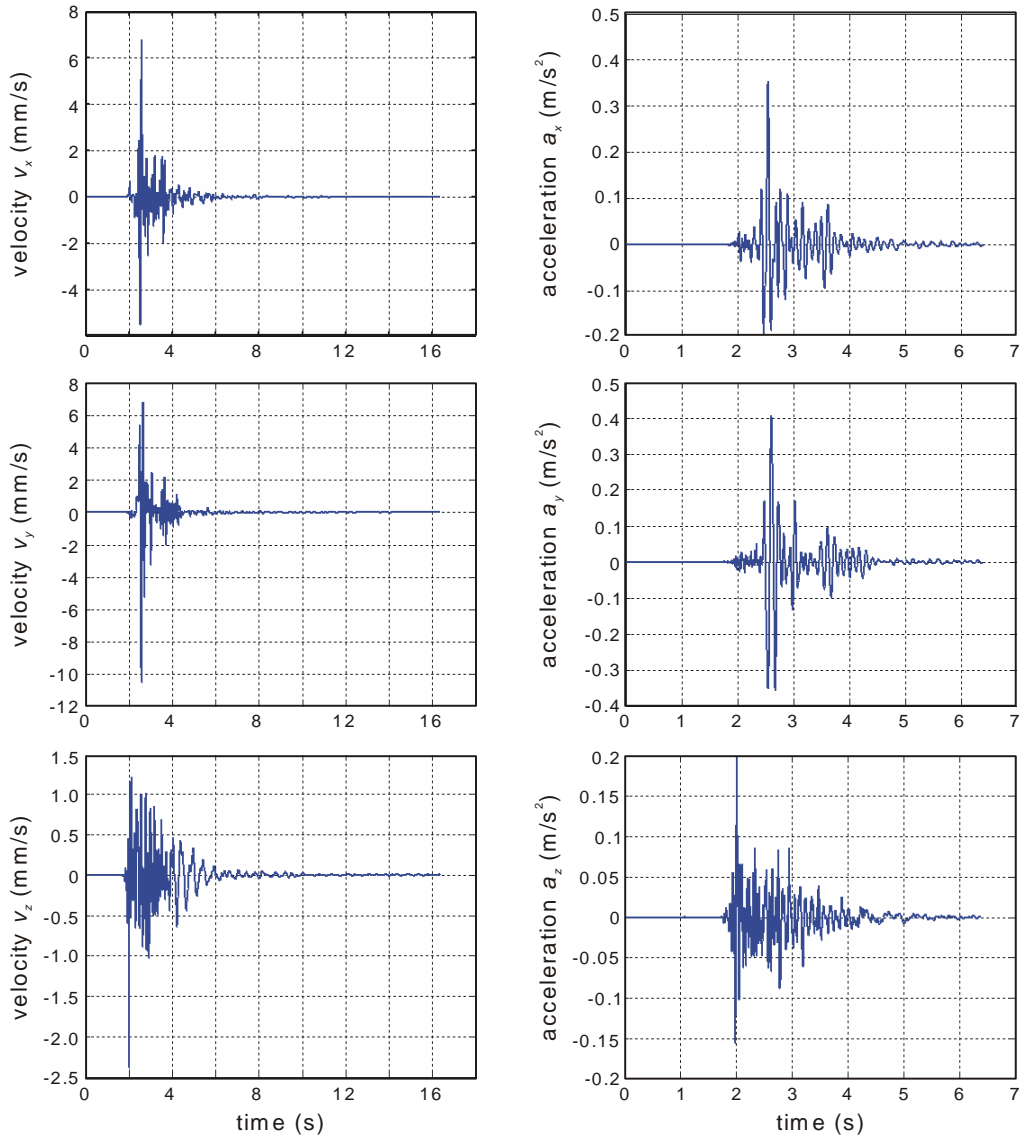

**Figure 11:** Components of the particle velocities (left) and accelerations (right) recorded at the basin center (point $F_2$).

*6.4.3 Numerical estimation of the ground motion in the basin*

Using the reference motion (Figure 10) and the numerical transfer function at the points located in the basin (Figure 8), the ground motion at the following 4 sites is computed: $F_2$, *A*, *B* and *C* (Figure 6). Denoting $S_{F_1}(\omega)$ the Fourier spectrum of the recorded motion at $F_1$ (location with no amplification), the spectrum $S_M(\omega)$ of the motion at any point *M* along the profile may be computed by using the transfer function $H_M(\omega)$ at this point in the following way:

$$S_M(\omega) = S_{F_1}(\omega) H_M(\omega) \tag{8}$$

First of all, it is checked that the acceleration computed at $F_2$ matches the recorded signals displayed in Figure 11. Combining the reference motion at $F_1$ (Figure 10) and the transfer function computed by the BEM at $F_2$ (Figure 8) for the *y* component of acceleration, the amplified acceleration at $F_2$ is obtained. As displayed in Figure 12, the time history of the computed acceleration (*y* component) is very close to the acceleration signal recorded at $F_2$ (Figure 11, center right). At point $F_2$, the maximum acceleration along *y* is found to be 0.407m/s$^2$ leading to an amplitude ratio of 2.63 when compared to the maximum acceleration in $F_1$. As usual, the time domain amplification is lower than the spectral amplification (Semblat and Pecker 2009).

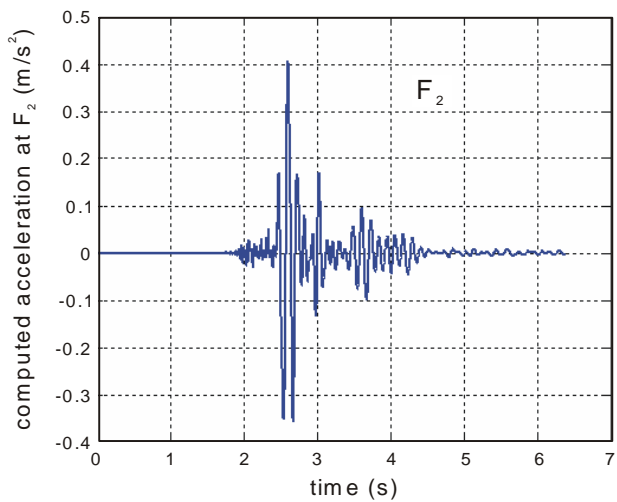

**Figure 12:** Acceleration at point $F_2$ computed from the reference signal at $F_1$ and the transfer function at $F_2$.

For points A, B and C, no recordings are available but similar computations are possible. The *y* component of acceleration is computed at each point and the Peak Ground Acceleration is compared to that obtained at point $F_1$. As shown in Table I, the amplification is around 35% at point A (0.210m/s$^2$ instead of 0.155m/s$^2$ at $F_1$). At point B, the acceleration is slightly larger than at A (amplification: 42%). At point C, the reference signal is amplified by around 20% only. The largest amplification (amplitude ratio: 2.63) is reached at point $F_2$, which is located over the deepest part of the deposit (Table I).

As already mentioned, the time domain amplification is lower than the spectral amplification. However, as shown by the transfer functions at different points (Figure 8), the amplification process is strongly influenced by the frequency content due to the depth variations in the deposit.

Finally, the amplification of the time signal is higher at points located over the alluvial deposit, which confirms the results obtained from the measurements (Driad-Lebeau et al., 2009), showing that the amplification is larger at points located in the vicinity of the center of the deposit.

**Table I**: Peak Ground Acceleration computed at different points and related amplifications.

| Locations | PGA (m/s$^2$) | Amplification/$F_1$ |
|---|---|---|
| Point $F_1$ | 0.155 |  |
| Point A | 0.210 | 1.35 |
| Point B | 0.220 | 1.42 |
| Point C | 0.190 | 1.23 |
| Point $F_2$ | 0.407 | 2.63 |

## 7  Conclusion

The propagation and amplification of mine induced vibrations was studied numerically by the Boundary Element Method. Surficial alluvial deposits are found to amplify the incident motion, which may lead to stronger excitations of the buildings located at the surface. A parametric study for various types of alluvial basins has been performed in order to make our numerical results applicable for different sites. The results provide the estimation of the amplification level and of the related frequency depending on both the shape ratio and the shear wave velocities ratio. Narrow basins lead to amplification levels and frequencies very different from the 1D case.

From the estimated geological profile of the Gardanne coal basin, the 2D amplification of the mine induced vibrations is then computed. These effects are found to significantly influence the ground motion at the free surface and thus the dynamic loading on the surface structures (i.e. buildings). For soft or deep surficial layers, the incident motion, even if moderate, may lead to significant ground motion at the surface of the deposit. Since they lead to different types of results, spectral and time domain amplifications must be studied simultaneously.

Future work will focus on the dynamic response of surface structures and the influence of the amplification of the ground motion.